# Pixel super-resolution interference pattern sensing via the aliasing effect for laser frequency metrology

*Lipeng Wan[1,2,3], Tianbao Yu[3], Daomu Zhao[2] and Wolfgang Löffler[1,*]*

1. Leiden Institute of Physics, Leiden University, 2333 CA Leiden, The Netherlands
2. Zhejiang Province Key Laboratory of Quantum Technology and Devices, School of Physics, Zhejiang University, Hangzhou 310027, China
3. School of Physics and Material Science, Nanchang University, Nanchang 330031, China

*Corresponding author:  loeffler@physics.leidenuniv.nl

**The superposition of several optical beams with large mutual angles results in sub-micrometer periodic patterns with a complex intensity, phase and polarization structure. For high-resolution imaging thereof, one often employs optical super-resolution methods such as scanning nano-particle imaging. Here, we report that by using a conventional arrayed image sensor in combination with 2D Fourier analysis, the periodicities of light fields much smaller than the pixel size can be resolved in a simple and compact setup, with a resolution far beyond the Nyquist limit set by the pixel size. We demonstrate the ability to resolve periodicities with spatial frequencies of ~3 μm$^{-1}$, 15 times higher than the pixel sampling frequency of 0.188 μm$^{-1}$. This is possible by analyzing high-quality Fourier aliases in the first Brillouin zone. In order to obtain the absolute spatial frequencies of the interference patterns, we show that simple rotation of the image sensor is sufficient, which modulates the effective pixel size and allows determination of the original Brillouin zone. Based on this method, we demonstrate wavelength sensing with a resolving power beyond 100,000 without any special equipment.**

## 1. INTRODUCTION

The precise measurement of periodicities of light fields is essential for various metrology tasks such as high precision laser frequency metrology, angular and position sensing, which is for instance crucial in nanolithography.[1] For investigation of complex field configurations, periodic patterns need to be analyzed, such as for the recently discovered bright superchiral fields that can be synthesized by interference of several optical beams,[2] and for exploration of more general interference phenomena.[3] Direct imaging of these fields (lens-free) is, however, considered to be impossible because of the large pixel size compared to the interference pattern periodicities, see Figure 1. The pixel size $\delta$ of modern CCD or CMOS image sensors is usually at least several micrometers due to limitations of the silicon base material and achievable signal to noise ratio,[4] resulting in a sampling frequency of $f_s = 1/\delta$. The Nyquist-Shannon sampling theorem [5,6] tells us that only structures with spatial frequencies smaller than $f_s/2$ can be resolved in all detail, leading to the condition $f_s/2 > f_L$, where $f_L$ is the spatial frequency of the interference pattern.

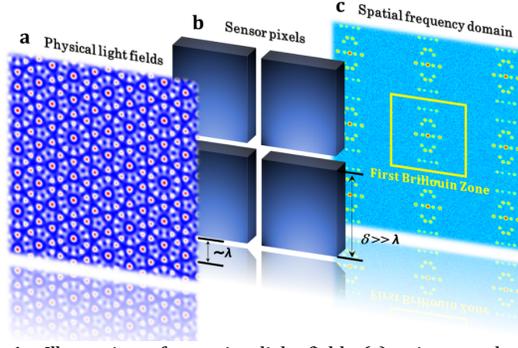

**Figure 1.** Illustration of mapping light fields (a) using standard image sensor whose pixel size (b) is much larger than the periodicities of the interference patterns (a). In the Fourier transform (c) of images captured by sensor, the aliases of high-spatial frequency components are clearly visible in the first Brillouin zone in the center as indicated by the yellow box spanned by the half-length reciprocal vectors.

Known methods for probing optical fields on scales much smaller than usual pixel sizes include nanoparticle scanning methods [7] near-field scanning optical microscopy (NSOM)[8,9] and vectorial field reconstruction [10]. The spatial resolution of these methods is ultimately limited by the size of probe to around 80 nm, and complex scanning equipment with nanometer precision is needed, and consequently these scanning-based methods are very slow.

Here we report that periodic light structures can reliably be detected with a simple arrayed image sensor such as CCD or CMOS by exploiting the aliasing effect happening for sampling well below the Nyquist limit, sketched in Figure 1. We found that 2D fast Fourier transforms are sufficient to determine the spatial frequencies of the aliases with high precision and that the relative phase of the beams can be retrieved unambiguously with high accuracy. The ambiguity of the measured spatial frequency of the aliases is resolved by changing the effective image sensor pixel size by rotating the image sensor. This method is very fast and the setup extremely simple since it only employs an image sensor mounted on a rotation stage and no imaging optics. We demonstrate measurement of interference periodicities 15 times smaller than the pixel size, beating the resolution of conventional pixel-based imaging. Using such a configuration, our resolution can, in principle, thousands of times higher than the image sensor resolution. A potential issue is the reduction of the signal-to-noise ratio (SNR) by undersampling,[11] but we find that this is not an issue here; another notable exception has been observed in experiments focusing light through opaque scattering media, where the SNR can be enhanced by 3–5 times with sub-Nyquist sampling. [12] Undersampling is often used in the temporal domain including in white-light interference microscopy as a function of path delay,[13] but no study has directly been explored in the spatial domain with a arrayed image sensor, to our best knowledge.

Exploiting this superresolution technique in the field of lensfree optical metrology, we demonstrate the prototype of a high precision wavemeter with a resolution of 5 pm using a 5.3 $\mu$m pixel-size image sensor. Reversely, if the laser wavelength is known, the same technique can be used to measure the angle between superposed dual laser beams with an accuracy of 8 $\mu$rad..

## 2. THEORY and ARRANGEMENT

Let us consider a superposition of $N$ plane electromagnetic waves with the same angular frequency $\omega = ck$ and fixed phase relation. The resulting electric field reads

$$\mathbf{E} = \mathrm{Re}\tilde{\mathbf{E}} = \mathrm{Re}\left(\sum_{j=1}^{N} \tilde{\mathbf{E}}_j e^{i(\mathbf{k}_j \cdot \mathbf{r} - \omega t)}\right). \quad (2)$$

The mean square of the electric fields is

$$S = \tfrac{1}{2}\tilde{\mathbf{E}} \cdot \tilde{\mathbf{E}}^* \\ = \tfrac{1}{2}\left(\sum_{l=1}^{N} \tilde{\mathbf{E}}_l \cdot \tilde{\mathbf{E}}_l^* + \sum_{j=1}^{N}\sum_{l \neq j} \tilde{\mathbf{E}}_j \cdot \tilde{\mathbf{E}}_l^* e^{i(\mathbf{k}_j - \mathbf{k}_l)\cdot \mathbf{r}}\right). \quad (3)$$

We obtain the corresponding field in the spatial-frequency domain by performing a 2D Fourier transform of the mean square of the electric field

$$\tilde{S}(\boldsymbol{f}) = \alpha \delta(\boldsymbol{f}) + \sum_{j=1}^{N}\sum_{l \neq j} \gamma_{jl} \delta(\mathbf{k}_j - \mathbf{k}_l + 2\pi\boldsymbol{f}), \quad (4)$$

where $\delta$ denotes the Dirac delta function, $\alpha$ is the zero-frequency component, and the complex weight function $\gamma_{jl}$ denotes the magnitude and phase of the spatial frequency component. We now discuss the superposition of two waves with wavevectors $k[\sin\psi, 0, \cos\psi]$ and $k[-\sin\psi, 0, \cos\psi]$; from which we obtain an interference pattern with spatial frequency $k\sin\psi/\pi$. We consider an arrayed image sensor with a pixel pitch of $\delta_x$ and $\delta_y$, which in our case for square pixels are equal, $\delta_x = \delta_y = \delta$. and the discrete Fourier transform of the light pattern sampled by the sensor array becomes (for the detailed derivation, see Supporting Information S1)

$$\tilde{S}_\mathrm{s}(f_x, f_y) = \sum_{u,v} \tilde{S}_\mathrm{p}\left(f_x - \tfrac{u}{\delta_x}, f_y - \tfrac{v}{\delta_y}\right), \quad (5)$$

where the complex function $\tilde{S}_\mathrm{p}$ gives the discrete response function

$$\tilde{S}_\mathrm{p}(f_x, f_y) = \tilde{S}(f_x, f_y) \cdot \tilde{P}(f_x, f_y). \quad (6)$$

Here $\tilde{P}$ is the Fourier transform of the two-dimensional pixel responsivity distribution, and the real integers $u$ and $v$ characterize the order of spatial aliasing. Equations (5) and (6) suggest that the recorded image on the image sensor can be represented as a set of reciprocal points in the spatial frequency domain (reciprocal space), each of which is described by the pixilated response $\tilde{S}_\mathrm{p}$. The analysis in the spatial frequency domain is performed within the first Brillouin zone or Wigner-Seitz cell spanned by the half-length reciprocal unit vectors. Physical insight into this process is gained by studying some cases. In the case that the spatial frequency of the light field is less than half the reciprocal vectors $f_L < (2 \cdot \delta)^{-1}$, there will be no overlap between different Fourier orders and thus, one can directly obtain the interference fringe periods of the superimposed light field. If this criterion, however, is not met, the high spatial frequencies will "fold back" into the first Brillouin zone leading to a frequency ambiguity, a phenomenon analogous to the "Umklapp process" in solid-state physics. In order to determine the original spatial frequency, we need to determine the original Brioullin zone number, for which we use rotation of the image sensor which can be modelled by a standard rotation matrix in Fourier space, say $\mathbf{R}_z(\theta)$. The spatial frequencies $\mathbf{f}_{jl}$ of the light field, generated by waves $j$ and $l$, are transformed into $\mathbf{f}_{jl}^r(\theta) = \mathbf{R}_z(\theta) \cdot \mathbf{f}_{jl}$, which means a change in the detected spatial frequencies because the effective image sensor pixel dimensions change, even for square pixels. We use continuous rotation and show that the absolute spatial frequencies can be retrieved in this way.

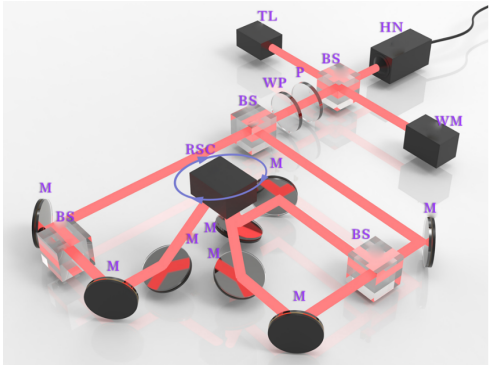

**Figure 2.** Experimental setup for pixel super-resolution interference pattern sensing: HN, He-Ne laser; TL, tunable laser diode NewFocus 6224; WM, wavemeter, HighFinesse WS6-200; RSC, rotating camera platform; M, mirror; BS, beam splitter; WP, half waveplate; P, polarizer.

Our experimental setup is shown in Figure 2. Although it is clear from Equation (6) that the pixel responsivity distribution of a specific image sensor is important, our method works for all types. Here we use a CMOS image sensor (Cinogy CMOS 1201-nano, i.e., a 5.3 μm pixel size image sensor without protective glass) mounted in the center of a precision rotation stage Newport M-URM80APP controlled by a Newport ESP300 controller, with angular resolution of 0.001 degree, allowing for continuous rotation of the sensor around its centered surface normal. In the process of retrieving the periodicities of the interference fringes of the superimposed light beams, all the necessary steps including rotation of the image sensor chip, recording of images and performing fast 2D Fourier transforms (FFTs) are automated by a computer.

## 3. MEASUREMENTS RESULTS

### A. One-dimensional interference patterns

We consider first two-beam superpositions. For this configuration, the electric field vectors are chosen to be $s$-polarized, $\tilde{\mathbf{E}}_{1,2} = [0, 1, 0]$ with respect to the image sensor plane, $\mathbf{k}_1 = k[\sin\psi, 0, \cos\psi]$ and $\mathbf{k}_2 = k[-\sin\psi, 0, \cos\psi]$ with $\psi = \pi/4$. From Equation (4) we determine that the spatial frequency of the interference pattern is 2234.85 mm$^{-1}$. The Nyquist frequency of our image sensor is 94.34 mm$^{-1}$, 23.7 times smaller. We record images while rotating the image sensor around its surface normal in steps of 0.1 degree. Interestingly, simply by observing the real-space image sensor images, we observed strong visual changes even for small rotation angles. We now perform real-time Fourier transforms and plot the result on a logarithmic color scale, as seen in Figure 3(a), where we observe next to the strong zero-frequency peak side-peaks that are aliases of the high-spatial frequency interference patterns as we show now.

We follow the Fourier peaks $f_m$ while rotating the image sensor, the result is shown in Figure 3(b). As expected for square image sensor pixels, the pattern shows a periodicity of 90 degrees, and mirror-symmetry with symmetry axes at $45 + n \cdot 90$ degrees, $n \in \mathbb{Z}$), and the good agreement between numerical calculation and experiment data is observed (For further information about numerical calculations and experiments, see Supporting Information S2). The modulation originates from the square shape of pixel, whose group of symmetries belong to $D_4$. However, it is still not straightforward to unravel the physical light field. To gain further insight into it, we perform the analysis over one cycle denoted by the orange box, followed by a decomposition into its Cartesian components, shown in Figs. 3(c) and 3(d). We observe an oscillation of the spatial frequency components upon rotation; a cycle is completed after rotation over 90 degrees. From the perspective of geometry, the projection of spatial frequencies of light fields on the x-axis should theoretically obey the relation $\mathbf{f} \cdot \hat{\mathbf{x}} = |f \sin(\pi\theta/180)|$ where $\hat{\mathbf{x}}$ is the $x$ unit vector. The observed oscillation pattern is in effect a manifestation of folding, where the high spatial frequencies beyond the first Brilliouin zone oscillate back and forth across the positive half of the first Brilliouin zone upon rotation. From this physical

picture, the x-component of true spatial frequencies $f_x^r$ upon rotation can be deduced to be

$$f_x^r(\theta, f_{bz}, f_x^m) = \left(m + \frac{1}{2} - \frac{1}{2}\cos m\pi\right) f_{bz} + f_x^m(\theta) \cos m\pi, \quad (7)$$

where $m$ denotes the $m$-th harmonics, $f_{bz} = f_s/2$ is the half-length of first Brillouin zone, and $\boldsymbol{f_x^m}$ is the measured x-component of the spatial frequency upon rotation. The y-frequency component $\boldsymbol{f_y^r}$ exhibits a reverse trajectory compared to the x-frequency component, due to the symmetry, and is

$$f_y^r(\theta', f_{bz}, f_y^m) = \left(m + \frac{1}{2} - \frac{1}{2}\cos m\pi\right) f_{bz} + f_y^m(\theta') \cos m\pi \quad (8)$$

where $\theta' = 90° - \theta$.

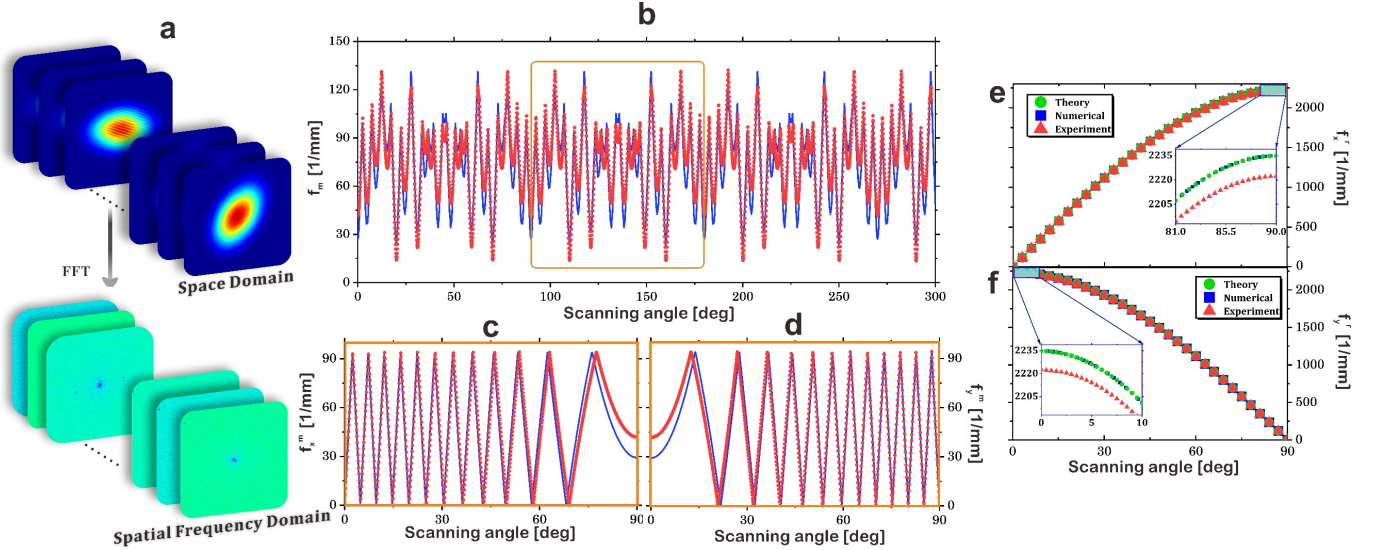

**Figure 3.** Probing superposed light fields using a CMOS image sensor with a 5.3 μm pixel pitch, where the structure of the interference pattern is much smaller than the pixel size. The image sensor is mounted on Newport M-URM80APP to form a custom-made rotating image sensor platform. For a continuous rotation (a) shows examples of captured intensity images in the space domain and the corresponding Fourier transform. Tracing the frequencies of Fourier peak gives (b) the measured trajectories of spatial frequencies; we compare numerical simulations (blue lines) to the experimental data (red dots). The orange box denotes one cycle. Within one cycle, we disentangle and extract the individual components of trajectories of spatial frequencies for the X-component (c) and Y-component (d). From this, we calculate the real spatial frequencies of the interference patterns for the X-component (e) and Y-component (f). The inset in (e) and (f) are magnifications of the spatial frequencies near the flat regions of the trajectories. The numerical and experimental data show good agreement.

Figures 3(e) and 3(f) present the genuine frequency trajectories of the physical light field retrieved using Equations (7) and (8), together with the corresponding theoretical prediction from projection. The image sensor rotation induces a change of the true spatial frequency projected along the axis, where the other frequency component disappears at the end of the rotation trajectory and thus, the true spatial frequency of the light field is retrieved. For the measurements of the X component, we retrieve the spatial frequency of physical light fields to be 2222.3759 mm$^{-1}$, in excellent agreement with the simulation and theory. The slight deviation in the true spatial frequency is attributed to the imperfect alignment of the light beams. This shows that we can simply measure the beam half angles with very high precision using our rotation technique, which is, in our case, 44.6811 degrees (the precision will be discussed later). Further, we observe that the retrieved spatial frequencies retrieved through the X- and Y-frequency components are in practice not equal, with the latter being 2222.7448 mm$^{-1}$. We argue that a residual tilt of the image sensor, or a small (75 pm) asymmetry of the image sensor pixels, is responsible for this.

**B. Probing superposition of multiple coherent beams of light**

To further demonstrate the power of our technique, we highlight that the proposed scheme can be applied to the general two-dimensional (2D) landscape as well. Light fields contain rich periodic features can be generated, for instance, by interfering multiple beams.[3,14] It is clear from Equation (3) that every beam-pair superposition $\widetilde{E}_j \cdot \widetilde{E}_l^*$ produces a pattern with high spatial frequency that contributes to the overall superposed light field distribution, we now explore the three waves configuration shown in Figure 2 and the specific parameters of this superposition are listed in Table 1, where $\psi = \pi/4$, $\varphi = \pi/4$, $\alpha$ and $\gamma$ are the relative phase of the waves. When rotating the image sensor, we observe three distinct Fourier peak trajectories, as expected. All these Fourier peaks are traced simultaneously during image sensor rotation, the frequency trajectories measured along the Y-direction are shown in

Figure 4(a), where the zone denoted by the orange box reveal a repeating pattern which are shifted for different beam combinations, containing information about their relative angles between the beams. By using Equation (8) one can retrieve the true spatial frequencies, whose retrieved trajectories are shown in Figs. 4(b). The true spatial frequencies can then be extracted from the end of the trajectories, where interference of the PR, PQ and QR beam combinations are, respectively, 1573, 2227 and 1577 mm$^{-1}$. The theoretical predictions for perfect alignment are 1580, 2235 and 1580 mm$^{-1}$, the errors are thus 0.45%, 0.36% and 0.19% which we attribute to a slightly misalignment of the wavevector. This indicates the high precision of our method, and that it is also applicable to multiple-beam superpositions.

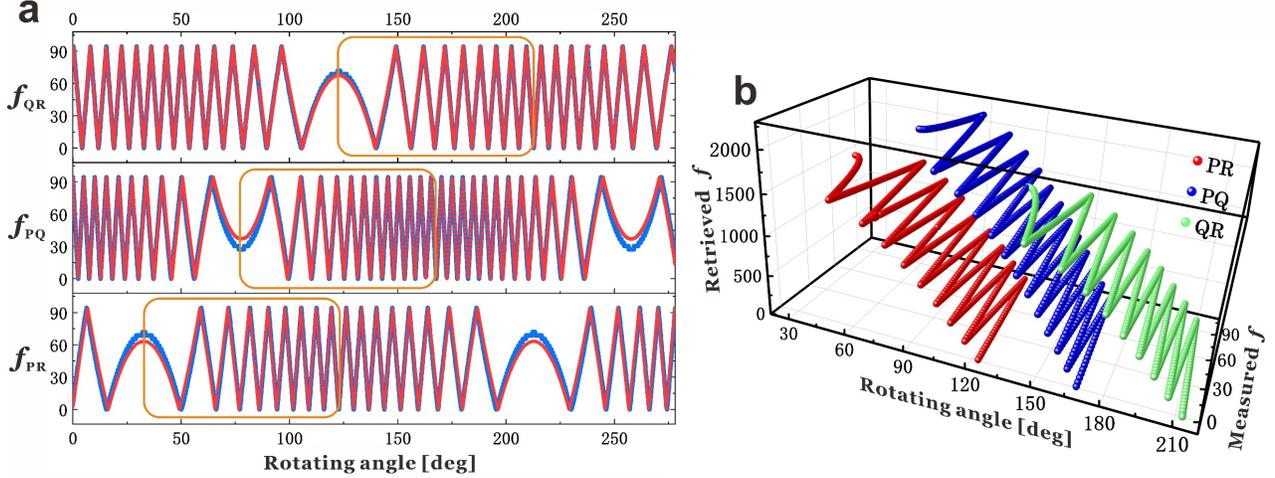

**Figure 4.** Resolving 3-beam light field superpositions. (a) Rotation measurements in the under-sampled case showing numerical simulations (blue) and experimental measurements (red). The orange box shows the repeating pattern that is angle-shifted for different beam combinations. (b) Retrieved spatial frequency trajectories for the 2D interference patterns. The spatial frequencies are shown in units of 1/mm.

**Table 1. 3-beam interference configuration**

| $j$ | $\mathbf{k}_j$ | $\widetilde{E}_j$ |
|---|---|---|
| P | [$\sin\psi$, 0, $\cos\psi$] | [-$\cos\psi\cos\varphi$, -$\sin\varphi$, $\sin\psi\cos\varphi$] |
| Q | [-$\sin\psi$, 0, $\cos\psi$] | [$\cos\psi\cos\varphi$, $\sin\varphi$, $\sin\psi\cos\varphi$]e$^{i\alpha}$ |
| R | [0, $\sin\psi$, $\cos\psi$] | [$\sin\varphi$, -$\cos\psi\cos\varphi$, $\sin\psi\cos\varphi$]e$^{i\gamma}$ |

## 4. PHASE RETRIEVAL

Having measured the spatial frequencies, we show that it is also possible to retrieve the relative phases of the beams in superposition using our method. From Equations (4)-(6) we see that the phase spectrum of Fourier transform is independent of the pixel pitch as well as two-dimensional pixel responsivity distribution, i.e., the precise properties of sensor therefore play no role for the phase spectrum of the Fourier transform. Therefore, the relative-phase information of the superposed light beams can be obtained, even in the undersampled case.

To prove this point, we attach a piezo chip (Thorlabs PA4HEW) to one of the mirrors in one of the arms, this allows for finely tuning the path difference with nanoscale precision, inducing a shift of the fringes of the interference pattern. We apply to the piezo a sawtooth signal with a frequency of 100 mHz, Figure 5(a) shows the measured phase of the Fourier peak. We see that the Fourier phase changes mostly linearly with the applied voltage. Synchronously, we measure the relative phase change of this Fourier peak compared to another Fourier peak originating from two beams without a piezo element, shown in Figure 5(b). We observe the same pattern except for a constant phase offset. This proves our theoretical predication, opening a new avenue to perform phase locking.

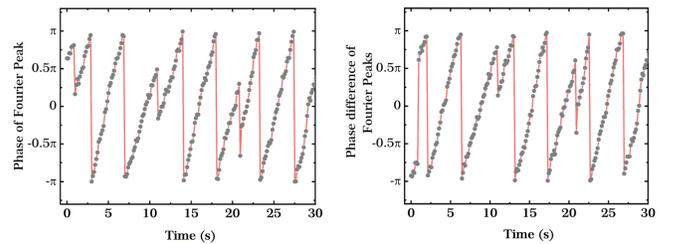

**Figure 5.** Demonstration of phase sensing. A sawtooth signal with a frequency of 100 mHz is imposed to the piezoelectric-actuated mirror. We measure the absolute Fourier peak phase (left) and the relative phase compared to another non-shifted beam combination (right).

## 5. SENSITIVITY AND WAVELENGTH SENSING

We now demonstrate application of our method to wavelength sensing.

*Theory.*—The maximum-magnitude FFT-pixel is taken for our spatial frequency reconstruction. The main inaccuracies of the setup are the precision of the rotation angle $\Delta\theta$, the precise image sensor dimensions, and a residual tilt of the image sensor. Based on the projected spatial frequency relation, we obtain for the wavelength resolution $\Delta\lambda$

$$|\Delta\lambda| = \frac{\lambda^2 \Delta f_c}{2\sin\psi + \lambda \Delta f_c}, \tag{9}$$

where $\Delta f_c$ is the spatial frequency resolution and is determined by the image size $L$ via $1/L$. This, in our case, yields an expected error of ~0.1 pm (~0.1 fm) for an angular rotation accuracy of $\Delta\theta$ = 0.1 deg (0.0001 deg). The finite size of the image sensor yields an error of 60 pm. The latter clearly appears to be the more important limit on resolution for our demonstration experiment. We can improve this by spatial interpolation via zero-padding, which enables a significant increase of the resolution of spatial frequency of the light. In our scheme, a zero-padding up to $2^{18} \times 2^{10}$ pixels is performed for all results. To ensure a good SNR, the interference size is relevant since it straightforwardly influences the width of the Fourier spots and thus, the accuracy of the measured spatial frequencies.

*Measurements.*—Experimentally, an external-cavity semiconductor laser at ~776.3 nm (New Focus model 6224) with a linewidth of less than 300 kHz is coupled into our setup via a single-mode fiber. We scan the wavelength and measure it using a Fizeau-interfero based wavemeter (HighFinesse WS6-200).

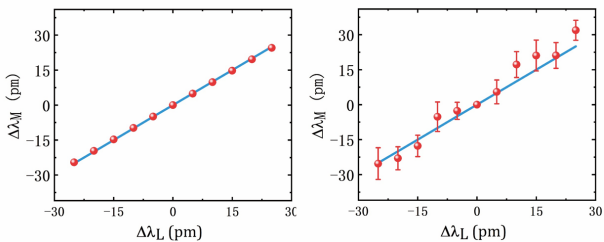

**Figure 6.** Demonstration of the high precision of wavelength sensing using our technique. The measured wavelength shift is indicated by red spheres for the numerical simulation (left) and experimental data (right). The blue line is theoretical prediction results. The error bars represent the standard deviations corresponding to 15 independent measurements.

Figure 6 shows our experimental results demonstrating ultra-high wavelength resolution. Due to the stability of the tunable laser source and the absolute accuracy of the wavemeter (200 MHz) the wavelength was scanned in steps of 5 pm. The observed linear relation is in good agreement with theoretical analysis and numerical calculation, thereby confirming the wavelength shifts down to at least 5 pm or 2.5 GHz in frequency are readily discernable by our method. This accuracy can easily be improved by enlarging the interference pattern region, increasing intensity, and using peak fitting for determination of the spatial frequencies of the Fourier aliases.

*Spectral performance.* We can derive the spectral resolving power $R$ that is equal to the ratio between the size of zero-padded image and the interference period (See Supporting Information S3). The simplicity of the inverse relation between the pixel size and the half length of the first Brillouin zone $\delta = 1/f_s$ means that once the aliasing order is established, a unique mapping of spatial frequencies to wavelength is possible, without the need of post-processing. For $\delta$ = 5.3 $\mu$m pixel size, this is equivalent to a free spectral range (FSR) of $\Delta\upsilon$ = 20 THz, much larger than the FSR of our Fizeau interferometers (WS6-200, 100 GHz) and standard Fabry-Perot cavities (several GHz). Our calculations in Equation (9) already resulted in a predicted wavelength accuracy below a femtometer, thus a spectral resolving power beyond $10^8$, bringing it to the required sensitivity for probing the Doppler wobbles induced by exoplanets [15] or the Zeeman-splitting of spectral lines of hydrogen and antihydrogen. [16] However, a direct proof of this theoretically achievable accuracy would require an ultrastable, high-SNR system.

From our simple setup it is evident that our technique could have several major advantages over the existing wavelength sensing techniques: higher resolution and experimental simplicity. Perhaps more importantly, systems containing optical elements such as lenses, gratings or glass blocks (Fizeau interferometers) cannot be used for high-energy photons where media are strongly absorbing. Our technique is therefore suited for the extreme-ultraviolet (EUV) and x-ray regime where recently ptychographic methods have been explored [17,18]– with the added benefit of experimental simplicity.

## 6. DISSCUSSION AND OUTLOOK

We have shown that a simple arrayed image sensor in combination with Fourier analysis allows for deep-subpixel sensing of periodic interference patterns. This is possible by exploiting aliasing, and the absolute spatial frequencies as well as phase spectrum can be obtained by rotation of the sensor, even if it is square.

In a proof-of-principle experiment we have demonstrated wavelength sensing with picometer resolution, potentially much simpler than Fizeau- or Fabry-Perot based interferometers. Our method can also quite easily be applied to all wavelength ranges where arrayed image detectors are available, for instance, to metrology challenges in EUV lithography.

Our approach is quite general, it can be applied to other wave systems in nature, ranging from electromagnetic and acoustic waves to matter waves. We emphasize that the remarkable accuracy can be further improved by enlarging the interference region, improvements on rotational accuracy.

A promising extension and initial motivation of the study is applying our methods to the observation and characterization of bright superchiral fields [2]. Additionally, combining our technique with concepts from quantum metrology could be interesting: For instance, in the case that interfering light fields are not coherent states of light but *N*00*N* states, an *N*-fold-enhanced resolution as compared with a classical interference lithography is possible, [19–21] using an image sensor sensitive to multi-photon absorption or with single photon resolution. This is reminiscent of recent schemes with pixel super-resolution quantum imaging, which have been achieved by measuring the joint probability distribution of the spatial resolution of spatially entangled photons.[22] Light fields with high-spatial frequency features appear also in other fields, such as by surface plasmon interference for nanolithography, at the interface between metals and dielectrics the wavelength of surface plasma waves can be down to the nanometer scale, while their frequencies remain in the optical range, going beyond the free-space diffraction limit of the light. [23–25]

**Acknowledgements:** The authors would like to thank Xinrui Wei, Jörg Götte, Koen van Kruining, and Robert Cameron for fruitful discussions, and Michel Orrit for providing the wavemeter. This research was supported by EU H2020 (QLUSTER, 862035) from NWO (QUAKE, 680.92.18.04), from NWO/OCW (Quantum Software Consortium), from the National Natural Science Foundation of China (NSFC) (11874321 and 12174338), and from the Fundamental Research Funds for the Central Universities (2018FZA3005).

**Conflict of Interest.** LW and WL filed a USPTO patent application "Image sensor-based interference fringe metrology with superresolution for laser beam angle and absolute frequency measurement" (number US 63/411252), which covers the concept and implementation described here.

**Data Availability Statement.** Data can be obtained from the authors upon reasonable request.

# Supporting Information

# Pixel super-resolution interference pattern sensing via the aliasing effect for laser frequency metrology

Lipeng Wan[1,2,3], Tianbao Yu[3], Daomu Zhao[2] and Wolfgang Löffler[1,*]

1. Leiden Institute of Physics, Leiden University, 2333 CA Leiden, The Netherlands
2. Zhejiang Province Key Laboratory of Quantum Technology and Device, School of Physics, Zhejiang University, Hangzhou 310027, China
3. School of Physics and Material Science, Nanchang University, Nanchang 330031, China

*Corresponding author: loeffler@physics.leidenuniv.nl

**Supplementary Information S1: Derivation of Eqs. (4)-(6)**

The Fourier transform of Eq. (3) is given by

$$\tilde{S}(f) = \int_{-\infty}^{+\infty} S(r) e^{-2\pi i f \cdot r} \, dr. \tag{S1}$$

On substituting Eq. (3) into Eq. (S1), one obtains the superposition of light fields in the spatial frequency domain, as described by Eq. (4) in the main text. The observation of light fields with the sensor $S_s(r)$ can be written as [1]

$$S_s(r) = [S(r) \circledast S_p(r)] \cdot \Gamma(r, \Delta r). \tag{S2}$$

Here the circled asterisk symbol is a 2D convolution of the superposed light fields with the pixel responsivity distribution [2], which is then filtered by a comb function $\Gamma$ consisting of an array of Dirac delta functions characterized by the pixel spacing $\Delta r = (\delta_x, \delta_y)$. Note that for CMOS sensor, the pixel spacing is, in general, not equal to the pixel size due to the imperfect fill factor.

We now consider the field in the spatial frequency domain $\tilde{S}_s(\mathbf{r})$. From the sifting property, the convolution with a Dirac delta function simply shifts its origin. By substituting Eq. (S2) into Eq. (S1) and further taken the convolution theorem into account (the Fourier transform of a convolution of two functions is equal to the product of the Fourier transforms of each function) we obtain

$$\tilde{S}_s(f_x, f_y) = \sum_{u,v} \tilde{S}\left(f_x - \frac{u}{\delta_x}, f_y - \frac{v}{\delta_y}\right) \tilde{P}\left(f_x - \frac{u}{\delta_x}, f_y - \frac{v}{\delta_y}\right). \tag{S3}$$

**Supplementary Information S2: Calculation of the undersampled light fields in the rotated frame and experimental data acquisition**

For the numerical calculation, we assume that the pixels are square. The sensor has 1280×1024 pixels with a pixel pitch of 5.3 $\mu$m. The calculation of intensity of superposed light fields is performed in subpixel units, which are incoherently added to yield the average light intensity on a sensor pixel. The sampling pitch of the light fields should be much smaller than the pixel pitch for an appropriate mapping the superimposed light field on the sensor, and is set to 0.081 μm in our calculations. The sensor noise, which is modeled by a white stochastic process (white Gaussian noise), is added to the intensity pattern. Note that the fill factor strongly affects the visibility $V$, but has no influence on the spatial frequencies of light fields (see Supplementary Information S4), thus unity fill factor is used in our model.

Experimentally, the light patterns were recorded by our CMOS image sensor. To ensure a good signal to noise ratio, our image sensor is set to automatically adjust to the optimal exposure time until the signal level is greater than 50% and stays in the range between 50% - 90% of full saturation level. Additional attenuation is used to avoid saturation. A single rotating scan with steps of 0.1 deg takes ~30 minutes for 1D configuration (over 90 degrees), in the 2D configuration requires a larger angle to be scanned and therefore takes longer, about 1 hour.

**Supplementary Information S3: Derivation of the spectral resolving power of our configuration**

The spectral resolving power is the capability to resolve two close-by wavelengths and is given by R = |λ/Δλ| with Δλ being the spectral resolution. Substituting the wavelength resolution given by Eq. (9) into $R$, one obtains the theoretical spectral resolving power of our setup

$$R = \frac{2\sin\theta}{\lambda \Delta f_c} + 1 = \frac{L}{D} + 1 \approx r \quad \text{(S4)}$$

where $L$ is the image size with zero padding, $D$ represents the spatial periodicity of the light fields and $r$ denotes their ratio $r = L/D$. Since we are interested in the case $L \gg D$, or say, $r \gg 1$, which implies that Nyquist–Shannon condition cannot be fulfilled and aliasing occurs. In this sense, the aliasing effect can be used to greatly improve the spectral resolving power.

**Supplementary Information S4: Influence of the camera fill factor**

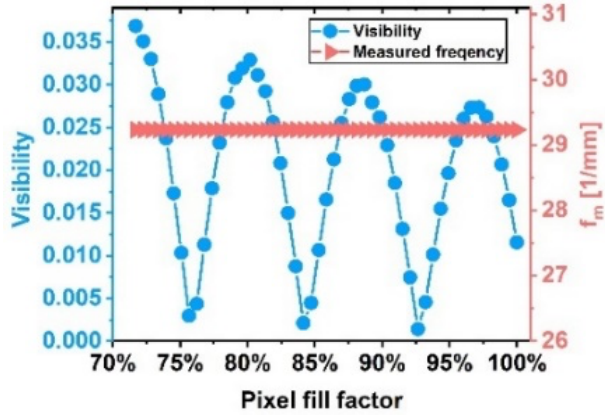

Fig. S1. Numerical experiments on the measurement for different fill factors. The measured spatial frequency (red markers, right axis) is uncorrelated to the pixel fill factor as expected, in contrast to the visibility of interference pattern (blue markers, left axis) which correlates closely with the pixel fill factor.

The influence of the fill factor of the pixel was investigated in a numerical experiment. Simulations were performed for square pixels with 10 nm subpixel units. The spacing between the pixels is varied from 0 to 1.5 $\mu$m in steps of 30 nm, corresponding a fill factor ranging from 71.7% to 100%. The other parameters are the same as for the other calculations. Note that CCDs often have a 100% fill factor but CMOS image sensors can have much less. The visibility and measured spatial frequencies are shown in Fig. S1. We observe that the interference visibility strongly depends on the fill factor, which can be seen as a manifestation of the fact that pixels do not sample point-like positions but sampling points are integrate over a larger area.